\documentclass[11pt]{article}
\usepackage[pdftex]{graphicx}
\usepackage{amssymb}
\usepackage{epstopdf}
\usepackage{multirow}

\usepackage[margin=1in]{geometry}

\usepackage[onehalfspacing]{setspace}

\newenvironment{affiliations}{%
    \setcounter{enumi}{1}%
    \setlength{\parindent}{0in}%
    \slshape\sloppy%
    \begin{list}{\upshape$^{\arabic{enumi}}$}{%
        \usecounter{enumi}%
        \setlength{\leftmargin}{0in}%
        \setlength{\topsep}{0in}%
        \setlength{\labelsep}{0in}%
        \setlength{\labelwidth}{0in}%
        \setlength{\listparindent}{0in}
        \setlength{\itemsep}{0ex}%
        \setlength{\parsep}{0in}%
        }
    }{\end{list}\par\vspace{11pt}}

\newcommand{\footlabel}[2]{%
    \addtocounter{footnote}{1}%
    \footnotetext[\thefootnote]{%
        \addtocounter{footnote}{-1}%
        \refstepcounter{footnote}\label{#1}%
        #2%
    }%
    $^{\ref{#1}}$%
}

\newcommand{\footref}[1]{%
    $^{\ref{#1}}$%
}

\title{ \textit{easyGWAS} : An integrated interspecies platform for performing genome-wide association studies}
\author{Dominik Grimm$^{1}$, Bastian Greshake$^{1}$, Stefan Kleeberger$^{1}$, Christoph Lippert$^{1}$,\\ 
Oliver Stegle$^{1}$, Bernhard Sch\"olkopf$^{2}$, Detlef Weigel$^{3}$ and Karsten Borgwardt$^{1,4}$}
\date{}
\begin{document}
\maketitle

\begin{affiliations}
\footnotesize
 \item Machine Learning and Computational Biology Research Group, Max Planck Institute for Developmental Biology and Max Planck Institute for Intelligent Systems, T\"ubingen, Germany
 \item Department of Empirical Inference, Max Planck Institute for Intelligent Systems, T\"ubingen, Germany
 \item Department of Molecular Biology, Max Planck Institute for Developmental Biology, T\"ubingen, Germany
 \item Center for Bioinformatics T\"ubingen, Eberhard Karls Universit\"at, T\"ubingen, Germany
\end{affiliations}

\tableofcontents
\newpage

\noindent\textbf{Abstract}

\noindent \textbf{Motivation:}
The rapid growth in genome-wide association studies (GWAS) in plants and
animals has brought about the need for a central resource that facilitates i)
performing GWAS, ii) accessing data and results of other GWAS, and iii)
enabling all users regardless of their background to exploit the latest statistical
techniques without having to manage complex software and computing
resources. 

\noindent \textbf{Results:} We present \textit{easyGWAS}, a web platform that provides methods, tools and dynamic visualizations to perform and analyze GWAS.
In addition, \textit{easyGWAS} makes it simple to reproduce results of others,
validate findings, and access larger sample sizes through merging of public
datasets.

\noindent\textbf{Availability:}
Detailed method and data descriptions as well as tutorials are available in the supplementary materials. \textit{easyGWAS} is available at http://easygwas.tuebingen.mpg.de/.

\noindent\textbf{Contact:} dominik.grimm@tuebingen.mpg.de

\section{Motivation}

Genome-wide association studies (GWAS) are an integral tool for discovering the polygenic architecture underlying many complex traits. The recent steady growth of GWAS applications across species (e.g. \textit{Arabidopsis thaliana} \cite{horton_genome-wide_2012,cao_whole-genome_2011,atwell_genome-wide_2010}, \textit{Drosophila melanogaster} \cite{mackay_drosophila_2012}) has generated a wealth of genotypic and phenotypic data, which makes it possible to search for significant association signals for multiple traits in one species or related traits in different species. Further analysing the shared associations across traits may provide invaluable insights for genetics and evolutionary biology: First, comparing results of GWAS in different species may enable statistical validation of association signals, for instance, to further support findings in human genetics by replicating compatible results in model organisms.  Second, one can examine the hypothesis that genetic factors influencing phenotypes can be traced back to related genes and mutations across species. By discovering these common genetic origins of phenotypes, we may gain a deeper understanding of the adaptation of species and possibly of the convergent or parallel evolution of complex traits.

\subsection{Difficulties in performing GWAS across traits and species}

Obtaining genome-wide association mapping results across multiple studies, traits or even species, however, is still a cumbersome enterprise, which is complicated by three problems: First, several software packages (e.g. PLINK \cite{purcell_plink:_2007}) or species-specific websites (e.g. DGRP \cite{mackay_drosophila_2012}, Matapax \cite{childs_matapax:_2012}, Emma-Server \cite{bennett_high-resolution_2010}) allow to perform genome-wide association studies on a given dataset. However, they either do not provide genotype and phenotype data at all or only for one single species. Second, existing databases for GWAS results (e.g. GWAS Catalog \cite{hindorff_potential_2009}) focus primarily on human genetics and present summary statistics only for the top scoring loci. This is despite the fact that the most significant genetic loci alone often explain only a small fraction of the heritability of complex traits \cite{manolio_finding_2009}. Third, databases for human genotypic and phenotypic data such as NCBI dbGaP \cite{mailman_ncbi_2007} require a formal application and evaluation before data access can be granted. Data from GWAS in model organisms and crops is more easily accessible in principle, but can only be obtained from individual websites in a variety of data formats. If one wants to run association studies with identical parameter settings on these datasets, one has to perform tedious data preprocessing and data integration steps first.

\subsection{Role of \textit{easyGWAS}}

The field is missing a platform that allows for easy and open access to published genotype and phenotype data from model organisms and crops and is able to perform GWAS on different traits and species. Here, we announce the release of \textit{easyGWAS}, an interactive and easy-to-use online platform, whose purpose is to fill this gap. \textit{easyGWAS} is available at http://easygwas.tuebingen.mpg.de/ and enables users to perform GWAS online on their own private or publically available data and to store and publish phenotypic data, meta information on the samples and GWAS results. 

\subsection{Functionality}

Genotypic and phenotypic data from a continuously growing set of published GWA studies are prestored in \textit{easyGWAS}. Users can either work with these public phenotypes or upload their own phenotypes.  Phenotype data uploaded by the user can either be kept private for primary analyses, shared with a restricted set of collaborators, or made publicly available to the community such that other researchers can reuse them in their GWAS analyses. \textit{easyGWAS} allows to perform univariate association tests in an interactive manner, without the need to manage any computing resources or software. Depending on whether the phenotype is binary or continuous, \textit{easyGWAS} offers suitable types of mapping algorithms to the user. The results of a completed GWAS can then be visualized in Manhattan plots with gene annotations for the top scoring signals. The example in section 3 and the Supplementary Material includes detailed instructions on how to perform GWAS in \textit{easyGWAS}. 

\subsection{Inter-compatibility with statistical genetics software packages}

If the user wants to perform an analysis, which is currently only available in existing statistical genetics software packages but not in \textit{easyGWAS} , the user can export \textit{easyGWAS} data and store them locally in a file format readable for these software packages. Data export to PLINK, comma-separated files (CSV) and hierarchical data format (HDF5) file format is already available in \textit{easyGWAS} .

\section{Example}
In this example of usage, we demonstrate how to perform a GWA study in the plant model organism \textit{Arabidopsis thaliana}. We use an already published phenotype~\textit{FLC}, which is related to the flowering time of the plant. For the \textit{FLC} phenotype, RNA was extracted from leaves after four weeks of growth and gene expression levels were determined by northern hybridization quantified relative to beta--tubulin expression.\\
Creating a new GWA experiment is divided into several intuitive steps. To be able to follow our instructions, the user should first navigate to the \textit{easyGWAS} experiment wizard by clicking on \textbf{GWAS Center} and then on \textbf{Create new GWAS}. 
\begin{enumerate}

\item First, the user selects a species and a dataset. In our example, we select the species \textit{Arabidopsis thaliana} and the dataset \textit{AtPolyDB (call method 75, Horton et al.)} and then click on \textbf{Continue}.

\item In the second step, the user has to select a phenotype. Here the user has the choice to select a \textbf{published}, \textbf{private/shared} or \textbf{public} phenotype. Published phenotypes are accompanied by a peer-reviewed research article, public phenotypes were made public by another user, but need not originate from a publication.
Additionally, the user can upload own data (see Supplementary Materials or the online FAQ for detailed tutorials). Here, we select an already published phenotype \textit{FLC} \cite{atwell_genome-wide_2010}. For this purpose, we select the tab \textbf{2.1 Select an existing published phenotype} and type into the input field the name of the phenotype \textit{FLC}. Auto-completion will support the user to select the correct phenotype. We proceed by clicking on \textbf{Continue}.

\item In this step, users can add additional factors such as principle components or covariates (e.g. environmental factors, gender specific characteristics). In our example, we do not add any additional factors. We click \textbf{Continue} in the tab \textbf{3.1 No additional factors}.

\item Now the user has to select genotypic data. Here, all provided SNPs, specific chromosomes or a region of SNPs can be selected. We select chromosome 1 and 5 in the tab \textbf{4.2 Select one or several chromosomes for \textit{Arabidopsis thaliana}} by checking the boxes and click on \textbf{Select chromosomes}.

\item To perform a GWAS, we have to select the association method we intend to use. In the algorithms view, one has options to also apply different transformations or filter to the data, such as normalizing the phenotypes. The selection of methods and transformations is dependent on the chosen data. Our web application is analyzing the data on the fly and is enabling only those options that are applicable for the chosen data. Here we keep the default settings using a \textbf{Linear Regression} without any transformations. Then we click on \textbf{Continue}.

\item In the last step, users can check all inputs and can make adjustments if necessary. If everything is correct, the experiment can be submitted to the computation servers. For this purpose, we simply click on \textbf{Submit Experiment}.

\end{enumerate}
Finally, the experiment is submitted and all computations are performed in the background. The current view refreshes every 3 seconds. In the meanwhile, users can submit new experiments or browse the data. Nevertheless, this example is finished in around 60 seconds and you will get automatically redirect to the result view to analyze the results. In the result view we provide dynamic Manhattan plots. Every single SNP can be explored in more detail by moving the mouse over a single point in the plot. On the left we provide a list with the top 10 SNPs and in which gene they are located. In our example the user can see at a glance, that for example the top three SNPs are located in chromosome 5. Additionally, we provide a more detailed SNP annotation view, quantile-quantile plots (QQ-plots) and a phenotype specific view with details about the phenotype (see Supplementary Materials for a detailed description and screenshots). Summary statistics can be downloaded in various formats for further analysis with third party tools.\\
Additional detailed tutorials (supported by screenshots) about uploading, sharing and downloading data are included in the Supplementary Materials.

\section{Conclusion and future plans}

\textit{easyGWAS} is designed to be a dynamically evolving platform with a growing number of functions and prestored datasets.  As of now, \textit{easyGWAS} offers published genotypic and phenotypic data for \textit{Arabidopsis thaliana} \cite{horton_genome-wide_2012,cao_whole-genome_2011,atwell_genome-wide_2010} and \textit{Drosophila melanogaster} \cite{mackay_drosophila_2012} and users can upload their own phenotypic data. \textit{easyGWAS} enables single-locus mapping with population structure correction for a single trait at a time. \\\\
In future versions of \textit{easyGWAS} , we plan to extend the list of species and to allow users to upload their own genotypic data, while retaining data quality and reliability. Further methods for multi-locus and multi-trait mapping and for automatically retrieving shared association signals across traits will be included.\\\\
In summary, we believe that \textit{easyGWAS} will foster new types of genetic analyses, by providing a convenient framework, which includes data and algorithms for obtaining GWAS results across traits, studies and species.

\newpage
\appendix

\section{Data: Genotypic \& phenotypic data and meta information}
\subsection{Available published data}
To easily perform genome-wide association studies (GWAS) across different species a variety of published genotypes and phenotypes are pre-stored in the \textit{easyGWAS} database. As of November 2012, data for \textit{Arabidopsis thaliana} and \textit{Drosophila melanogaster} are available in our public database. For \textit{Arabidopsis thaliana} we included different data sources. The first dataset ['AtPolyDB (call method 75, Horton \textit{et al.})'] includes 1,307 samples presented by Horton \textit{et al.} in 2012 \cite{horton_genome-wide_2012}. Furthermore, we included 107 phenotypes, described and analyzed by Atwell \textit{et al.} \cite{atwell_genome-wide_2010}. These 107 phenotypes are measured for a subset of these 1,307 samples. The second dataset ['80 genomes data (Cao \textit{et al.})'] includes 80 samples from the first phase of the 1001 genomes project in \textit{Arabidopsis thaliana} \cite{cao_whole-genome_2011}. The genome matrix from the 1001 genomes website\footnote{www.1001genomes.org} was used to retrieve all single nucleotide polymorphisms (SNPs). For this purpose, we excluded all positions with incomplete information and kept all positions with at least one consecutive nucleotide. All SNPs in these \textit{Arabidopsis thaliana} datasets are homozygous ones. Each allele in the SNPs is encoded as described in Table \ref{tab:alleleencoding}.
\begin{table}[h]
\begin{center}
    \begin{tabular}{|c|c|c|}
    \hline
     & major allele & minor allele \\ \hline
    major allele & 0 & 1  \\ \hline
    minor allele & 1 & 2 \\ \hline
    \end{tabular}
\end{center}
\caption{SNP encoding}
\label{tab:alleleencoding}
\end{table}

For the species \textit{Drosophila melanogaster} we integrated a dataset ['Drosophila Genetic Reference Panel (DGRP, Mackay \textit{et al.})'] with 172 samples \cite{mackay_drosophila_2012}, sequenced and analyzed by Mackay \textit{et al.}, as well as three phenotypes \cite{mackay_drosophila_2012, harbison_quantitative_2004, morgan_quantitative_2006, jordan_quantitative_2007} (six phenotypes, after splitting those into male and female). Missing SNPs in the \textit{Drosophila melanogaster} genome were imputed using the majority allele (different modes of imputation are currently being included into  \textit{easyGWAS}  and will be available soon).\\
Additionally we integrated gene annotations for all organisms. This information is used to identify if a SNP is located within a gene or not.\\
Publicly available genotypes and phenotypes are accompanied by additional meta information such as growth conditions in \textit{Arabidopsis thaliana} or \textit{wolbachia status} in \textit{Drosophila melanogaster}.
All datasets were downloaded from their official websites (Table \ref{tab:datasources}).

\begin{table}[h]
\hspace{-0.4cm}
\footnotesize
   \begin{tabular}{|l|l|l|}
\hline
 \multirow{3}{*}{AtPolyDB (call method 75, Horton \textit{et al.})} & Genotypes & https://cynin.gmi.oeaw.ac.at/home/resources/atpolydb \\
 &  \multirow{2}{*}{Phenotypes} &  https://cynin.gmi.oeaw.ac.at/home/resources/atpolydb\\
 & & http://arabidopsis.gmi.oeaw.ac.at:5000/DisplayResults/ \\\hline
 80 genomes data (Cao \textit{et al.}) & Genotypes & http://1001genomes.org/data/MPI/MPICao2010/releases/ \\\hline
 \textit{Arabidopsis thaliana} annotations & Annotations & http://www.arabidopsis.org/ \\\hline
  Drosophila Genetic Reference  & \multirow{2}{*}{Genotypes} & http://dgrp.gnets.ncsu.edu/freeze1/Illumina \\
  Panel (DGRP, Mackay \textit{et al.}) &  &  \_+\_454\_SNP\_genotypes\_filtered\_for\_GWAS/\\
   & Phenotypes &  http://dgrp.gnets.ncsu.edu/freeze1/Phenotypes/\\\hline
 \textit{Drosophila melanogaster} annotations & Annotations & ftp://ftp.flybase.net/releases/FB2008\_10/dmel\_r5.13/gff/ \\\hline
\end{tabular}
\caption{Data sources for all integrated organisms}
\label{tab:datasources}
\end{table}

\subsection{How to upload new data}
\subsubsection{Phenotypic data and meta information}
Registered users can upload private data such as phenotypes, covariates or meta information using the \textit{easyGWAS} wizard (see \textit{Tutorial 5.2}). This data can be used to perform new GWAS or can be shared with collaborators and colleagues. Furthermore, data can made be public to the scientific community. We distinguish between \textbf{published} and \textbf{public} data. Published data is integrated by the \textit{easyGWAS} team using data from peer-reviewed publications. However, \textbf{public} data was made public by any \textit{easyGWAS} user.  We also provide a contact form for authors who would like to have their published phenotype data integrated into {\it easyGWAS} through the \textit{easyGWAS} team, rather than uploading their data themselves.

\subsubsection{New genotypic data}
\begin{figure}[h] 
   \centering
   \includegraphics[width=12cm]{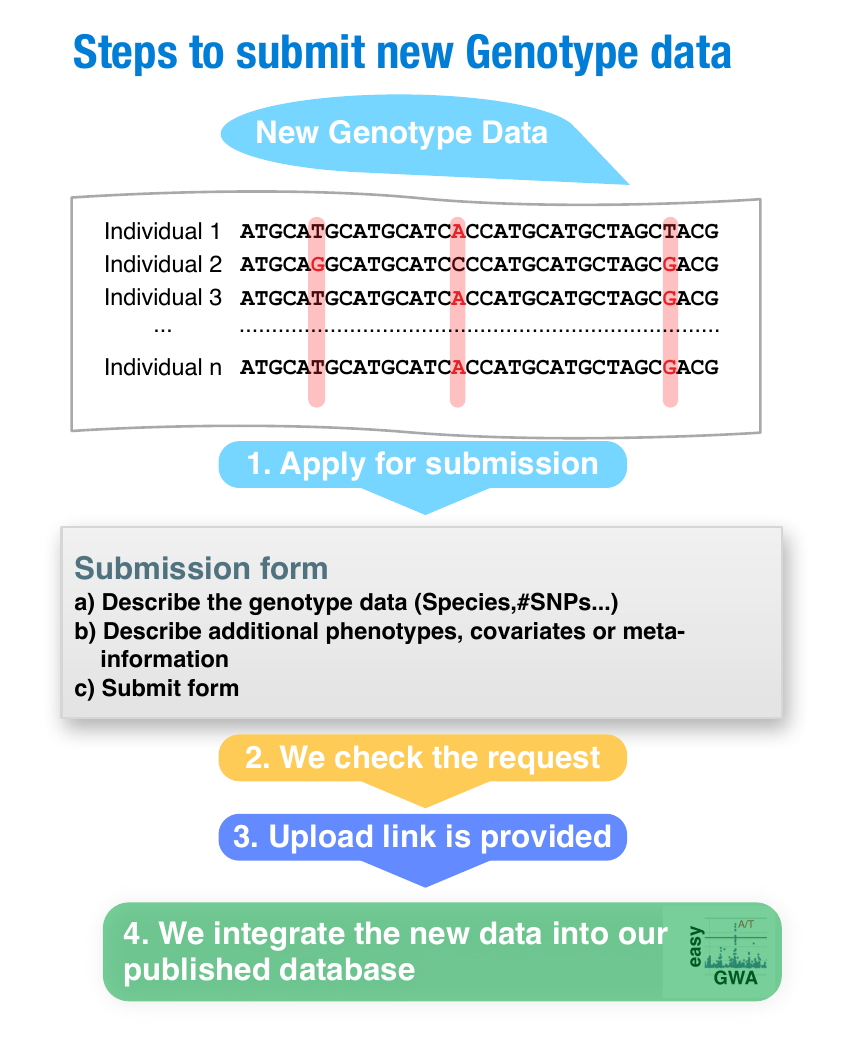} 
   \caption{Application process to submit new genotypic data}
   \label{fig:dataapply}
\end{figure}

Until now, we provide different datasets for two species. We plan to include datasets from different species to provide a richer selection of genotypic and phenotypic data. To retain quality, we provide an application form which users can use to send a formal data submission application to us. We then will evaluate the request. After successful evaluation, we will provide an upload link and after successful upload and quality inspection of the data, our team will include the data into our database (Figure \ref{fig:dataapply}).\\
A future extension will be a private upload option for small genotypic data sets.  

\section{Integrated methods to perform genome-wide association studies}
In general, performing a genome-wide association study is not trivial. There are three main aspects that have to be considered. The first category is data preprocessing. The scientist has to know how to encode, normalize and filter the genotypes, phenotypes and covariates. Second, the scientist has to know which method can be used for which kind of data. There are binary and continuous phenotypes as well as homozygous and heterozygous genotypes. Only specific methods can be applied to a specific type of data. Third, it is crucial to decide whether one should correct for population stratification or latent confounding factors. Further, some of these methods are hard to parameterize or complicated to set up. One of the strengths of  \textit{easyGWAS}  is that it provides several implemented methods and data transformations out of the box. This helps the user to easily perform a genome-wide association study.

\subsection{Methods to perform a GWAS}
The initial version provides several univariate algorithms, such as linear regression, linear mixed models (EMMAX \cite{kang_variance_2010}, FaSTLMM \cite{lippert_fast_2011}) and the Wilcoxon rank-sum test.  Linear regression can be used to find single associations between a single SNP and a phenotype. Linear mixed models are used to correct for population structure, family structure and cryptic relatedness at the same time. To all these methods one can add covariates, such as principle components (PCs), environmental factors or gender specific characteristics. Additionally, the Wilcoxon rank-sum test can be used for homozygous genotypes. These methods are state-of-the-art, more methods will be added continuously. New methods for multi-marker discovery such as two-locus search using graphical computing units (GPUs) and multi-trait discovery will be added in the near future.

\subsection{Transformations to standardize data}
\begin{figure}[h] 
   \centering
   \includegraphics[width=17cm]{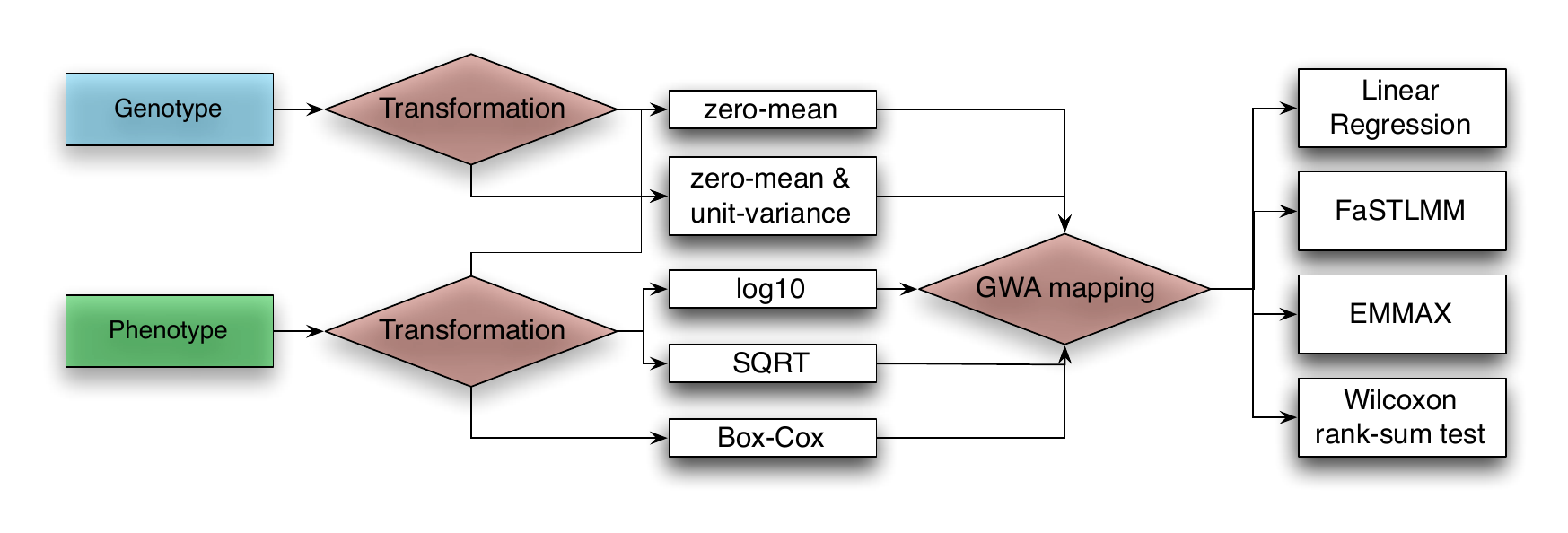} 
   \caption{Scheme off all possible transformations and GWA mapping methods}
   \label{fig:algo}
\end{figure}

To transform phenotypic and genotypic data we added several methods. Genotypes can be standardized, one can zero-mean the data and/or divide by unit variance. Phenotypes can be transformed in the same way. Additionally, phenotypes can be log-transformed, square root and box-cox transformed. Figure \ref{fig:algo} illustrates a scheme of all options.

\section{The web application interface}
The web application contains three main parts. There is one view to plan, perform and store GWAS, a second view to browse and analyze the data and a third view to download available datasets. In the following, we will describe all sections in detail.

\subsection{The \textit{easyGWAS} wizard and experiment history}
\begin{figure}[h] 
   \centering
   \includegraphics[]{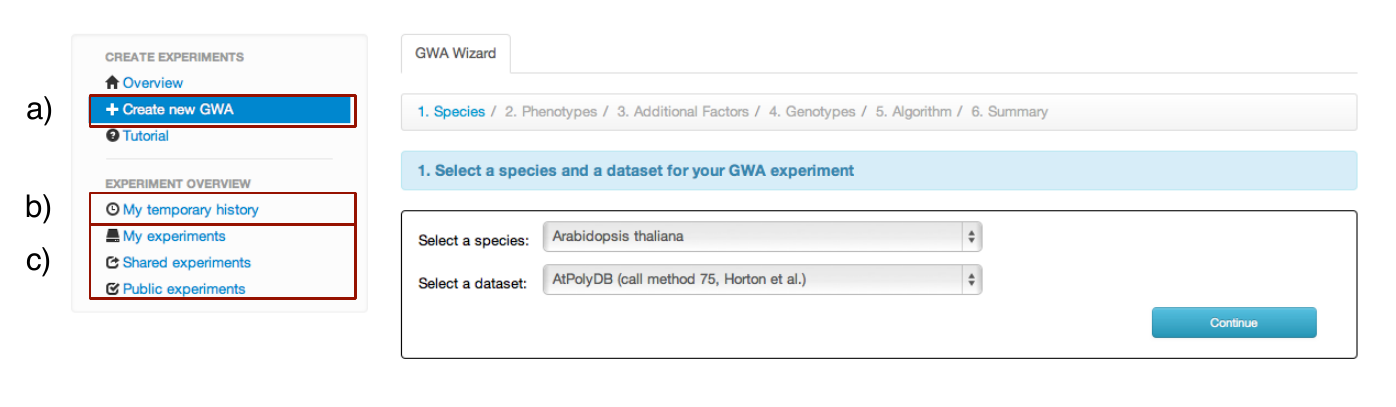} 
   \caption{Screenshot from the GWA experiment view. Creating new GWA experiments and sharing the results with collaborators or the scientific community}
   \label{fig:wizard}
\end{figure}

\begin{figure}[h] 
   \centering
   \includegraphics[]{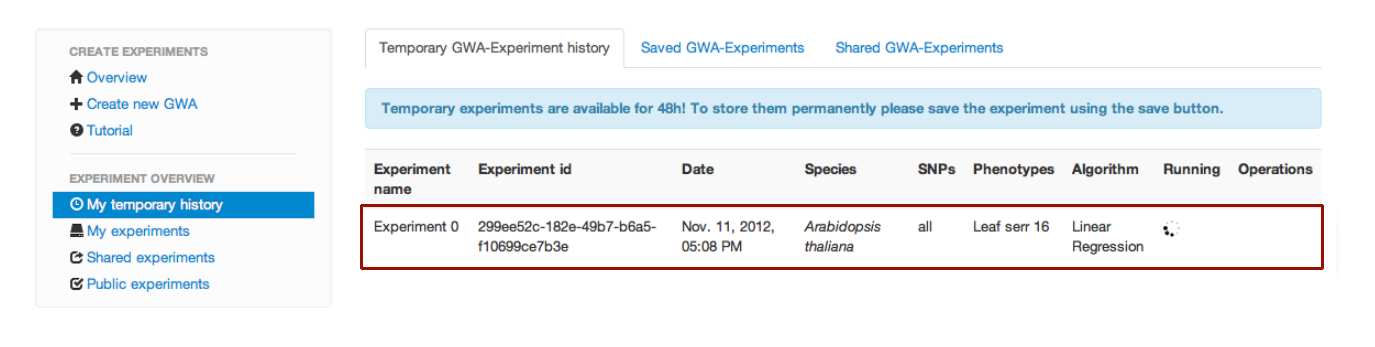} 
   \caption{Screenshot from temporary experiment history. The red highlighted row indicates that the experiment is still computing.}
   \label{fig:running}
\end{figure}

\begin{figure}[h!] 
   \centering
   \includegraphics[]{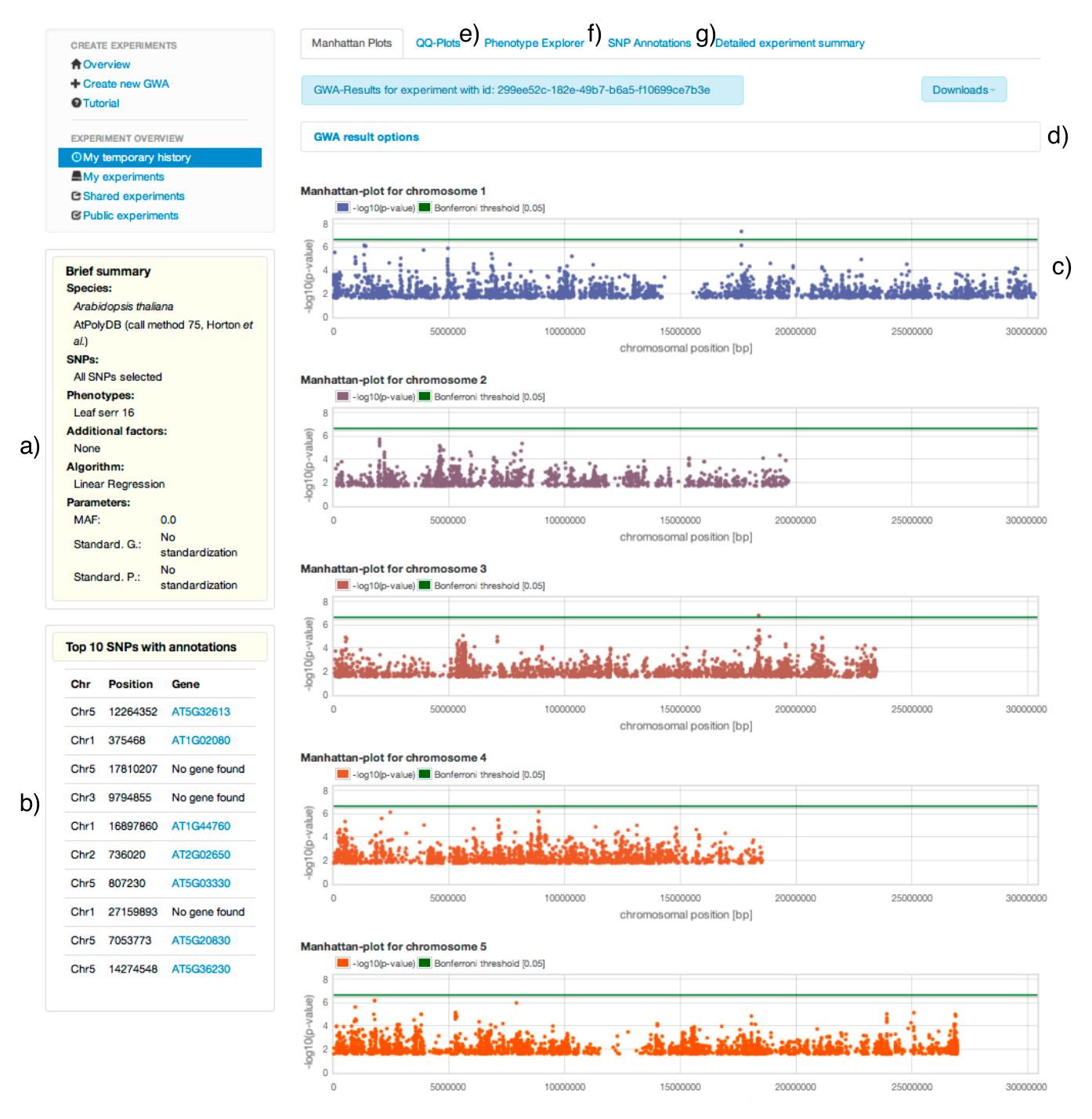} 
   \caption{Screenshot showing the result page of an experiment.}
   \label{fig:resultpage}
\end{figure}

The first section contains all necessary tools to plan, perform and analyze whole genome-wide association studies. Here registered users can use a step-by-step procedure (software wizard) to easily create new experiments (see Tutorial 5.1, Figure \ref{fig:wizard}a). 
The wizard is divided into several steps. First the user has to choose an available species and dataset. In the next step a single phenotype can be selected. Here it is possible to select an already published, private or public phenotype. Additionally, one can upload an own phenotype. We distinguish between {\bf published} and {\bf public} phenotypes. {\bf Published} phenotypes are already published, whereas {\bf public} phenotypes are phenotypes uploaded by a specific user and made public to the community. After chosen a phenotype it is possible to add additional factors to the experiment, such as principle components or one or several covariates. Covariates are meta information such as environmental factors or gender specific characteristics. To proceed the user has to select genotypic data. To do so, it is possible to select all genotypic data, meaning all available SNPs. Furthermore, specific chromosomes or a range of SNPs can be selected. The last step provides different algorithms, standardizations and filters. The selection of methods is based on the selected genotypic and phenotypic data in the previous steps. The summary view in the end provides all user specific selections and offers the user to adjust settings or to submit the experiment to the computation workers. \\
Each experiment performed is saved in a temporary experiment history (Figure \ref{fig:wizard}b). Here all experiments are stored for primary analysis for at least 48h. To keep interesting findings, users can store experiments permanently in their private profile. To simplify scientific exchange, all experiments can be shared via the web application with collaborators and colleagues (Figure \ref{fig:wizard}c). Sharing experiments and data can prevent laborious extracting of data and findings. Furthermore, data and experiments can be made public to the scientific community. All summary statistics can be downloaded to further analyze the data using third party tools.\\
Performing GWAS can be time consuming. Due to an advanced technology (see \textit{Web application infrastructure}) the user can continue working using the web application while the experiment is computed in the background at the same time. Automated email notifications are send out as soon the computations are done. Additionally, the user can track the status of all experiments through the temporary history (Figure \ref{fig:running}).
\\
To examine individual experiments each experiment has an interactive results page. Figure \ref{fig:resultpage} shows a screenshot of the result page. The view is divided into two parts. The left part provides general information. Here a short summary table informs about all settings made by the user, e.g. which species, dataset and parameters were selected (Figure \ref{fig:resultpage}a). 
At a glance the user can see the top 10 SNP annotations with the smallest p-values (Figure \ref{fig:resultpage}b). Dynamic Manhattan-plots for all chromosomes are rendered in the right half (Figure \ref{fig:resultpage}c). Each SNP within the Manhattan plot is interactive, meaning that the user is able to inspect single SNPs getting live information like the corresponding p-value or in which gene the SNP is located. The green line in each Manhattan-plot is the Bonferroni threshold. The alpha significance level can be adjusted using the plotting options (Figure \ref{fig:resultpage}d). The strength of population structure confounding can be easily explored with Q-Q plots (Figure \ref{fig:resultpage}e) and the genomic control $\lambda$. To see the actual distribution of a phenotype, the \textit{Phenotype Explorer} shows histograms for transformed and non-transformed phenotypes and computes a Shapiro-Wilk test to test the null hypothesis that the data was drawn from a normal distribution (Figure \ref{fig:resultpage}f). To examine if SNPs of interest are located within genes, significant loci are summarized in a gene-annotation view (Figure \ref{fig:resultpage}g). 

\subsection{The data center}
\begin{figure}[h!] 
   \centering
   \includegraphics[]{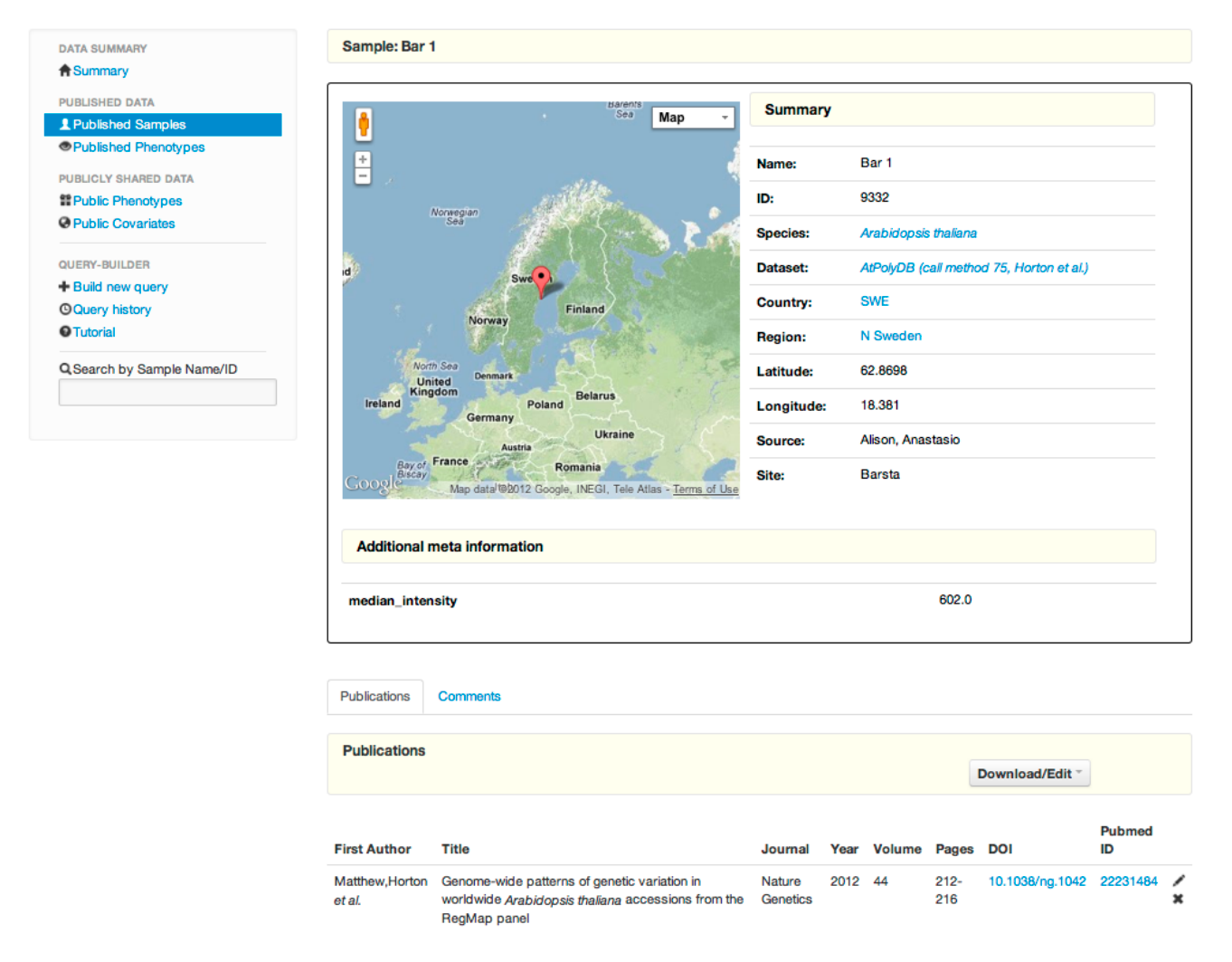} 
   \caption{Data center with detailed information about sample, phenotypes and covariates}
   \label{fig:sampleview}
\end{figure}

The second main section of the web application is the \textbf{Data Center}. Here, the user can browse available data, such as samples, phenotypes and covariates. Detailed information can be accessed for each data entry, such as meta and/or geographical information (Figure \ref{fig:sampleview}). Associated publications are provided for all published entries. The \textbf{Data Center} contains two main views. One for \textbf{published} and \textbf{public} data and a second for all user specific \textbf{private/shared} data. Private data is only visible to the owner of the data. Note that privately shared data belongs to the owner, meaning that only the owner has the permission to delete or modify shared data.
\\

\subsection{The download center}
The third section provides additional download options. Here, whole datasets (genotypic and phenotypic data) for all integrated species can be downloaded in different file formats. At the moment we provide the following formats: PLINK\cite{purcell_plink:_2007}, comma-separated files (CSV)  and hierarchical data format 5 (HDF5)\footlabel{hdf5}{http://www.hdfgroup.org/HDF5/}. 

\section{The web application backend}
The backend of the web application is completely written in Django\footnote{https://www.djangoproject.com} a web framework for Python. For the web design, we used the Cascading Style Sheets (CSS), provided by Twitter Bootstrap\footnote{http://twitter.github.com/bootstrap/}. 
\begin{figure}[h] 
   \centering
   \includegraphics[]{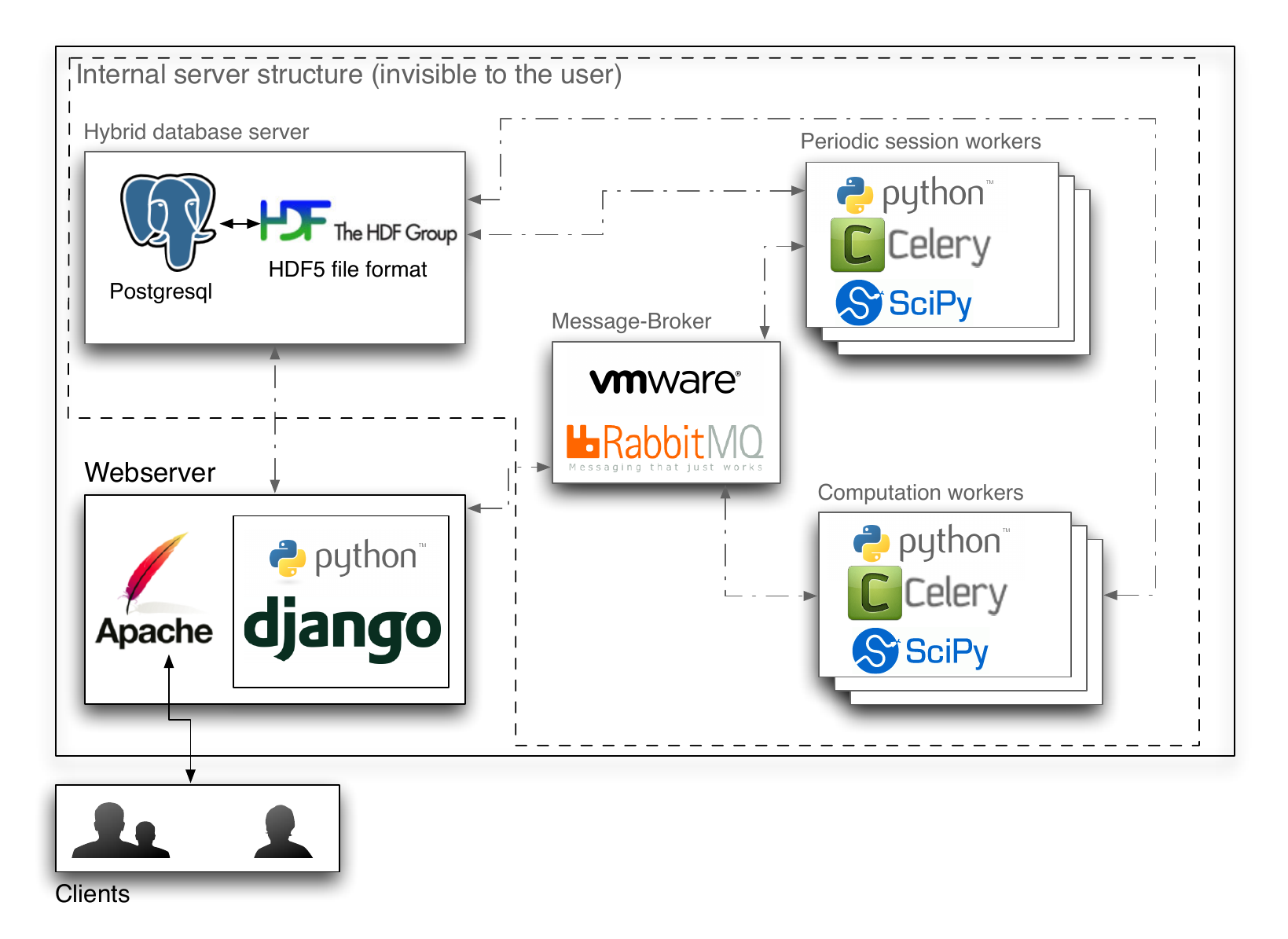} 
   \caption{Scheme of the web application infrastructure}
   \label{fig:webinfra}
\end{figure}
To handle the huge amount of SNP data we developed a hybrid database model using a PostgreSQL database and the HDF5\footref{hdf5} file format. Here all SNPs, associated positions and chromosome indices as well as all phenotypes and covariates are stored in the HDF5 file. Additional phenotype, sample, covariate and meta information are stored in the PostgreSQL database. HDF5  files are highly optimized to handle huge files and can be accessed fast and easily. As GWA mappings are resource consuming all computations are distributed to different computation servers (workers). To schedule different tasks smartly we are using a message broker (RabbitMQ\footnote{http://www.rabbitmq.com}). This broker distributes the different tasks to single workers (Figure \ref{fig:webinfra}). The backend is well designed to easily extend the functionality of \textit{easyGWAS} using additional novel methods. New species and datasets can be integrated within hours, depending on the size of the data. If more computational power is needed, new workers can be added dynamically.

\section{Tutorials}
In this section we provide various tutorials on how to use \textit{easyGWAS}. We demonstrate how to actually perform a genome-wide association mapping, how to upload own private phenotypes and how to share or make them public for collaborators and/or the scientific community. Furthermore, we show how to download summary statistics of GWA experiments and published genotype and phenotype data. Screenshots are attached to all important steps.

\subsection{How to perform a GWAS easily?}
In this tutorial we demonstrate how to easily perform a GWA study using \textit{Arabidopsis thaliana} and already published phenotype \textit{FLC} (flowering time related phenotype).
\begin{enumerate}
	\item If not already done: Create a new \textit{easyGWAS} account and log in.
	
	\item Navigate to the \textit{easyGWAS} wizard\\
		\textbf{Menu: GWA-Experiments $\rightarrow$ Create new GWA}
	
	\item First, select a species and a dataset. Here we choose the species \textit{Arabidopsis thaliana} and the dataset \textit{AtPolyDB (call method 75, Horton et al.)}\cite{horton_genome-wide_2012} and click \textbf{Continue}.
		\begin{center}
		  \includegraphics[]{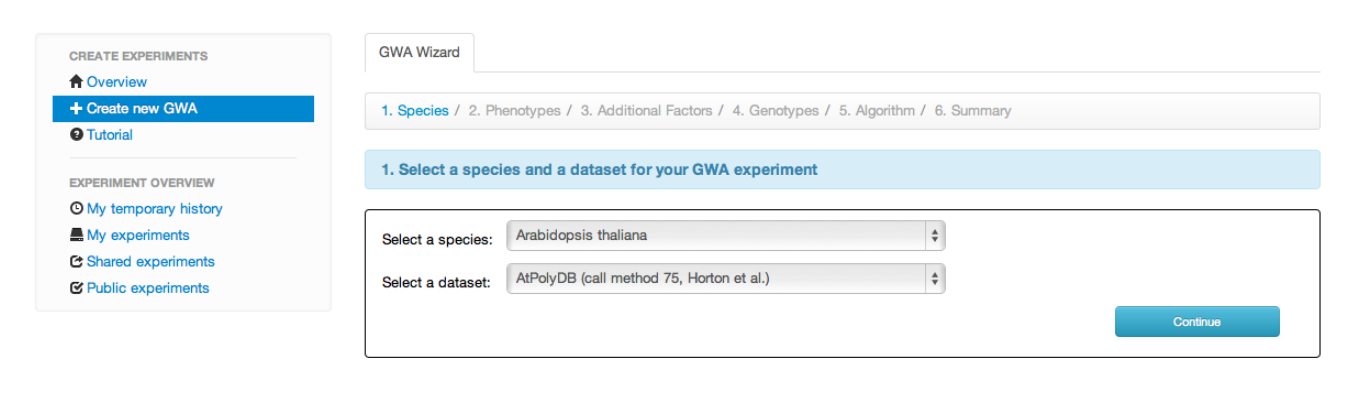} 
		\end{center}
	
	\item Select a phenotype. Here the user has the choice to select published, private/shared and public phenotypes. Additionally the user can upload his own data (see Tutorial 5.2). Here we select a published phenotype \textit{FLC}\cite{atwell_genome-wide_2010}. For this purpose, select the tab \textbf{2.1 Select an existing published phenotype} and type into the input field the name of the phenotype (\textit{FLC}\cite{atwell_genome-wide_2010}). Auto-completion will help you to select the correct one. Click \textbf{Continue}.
	\begin{center}
		  \includegraphics[]{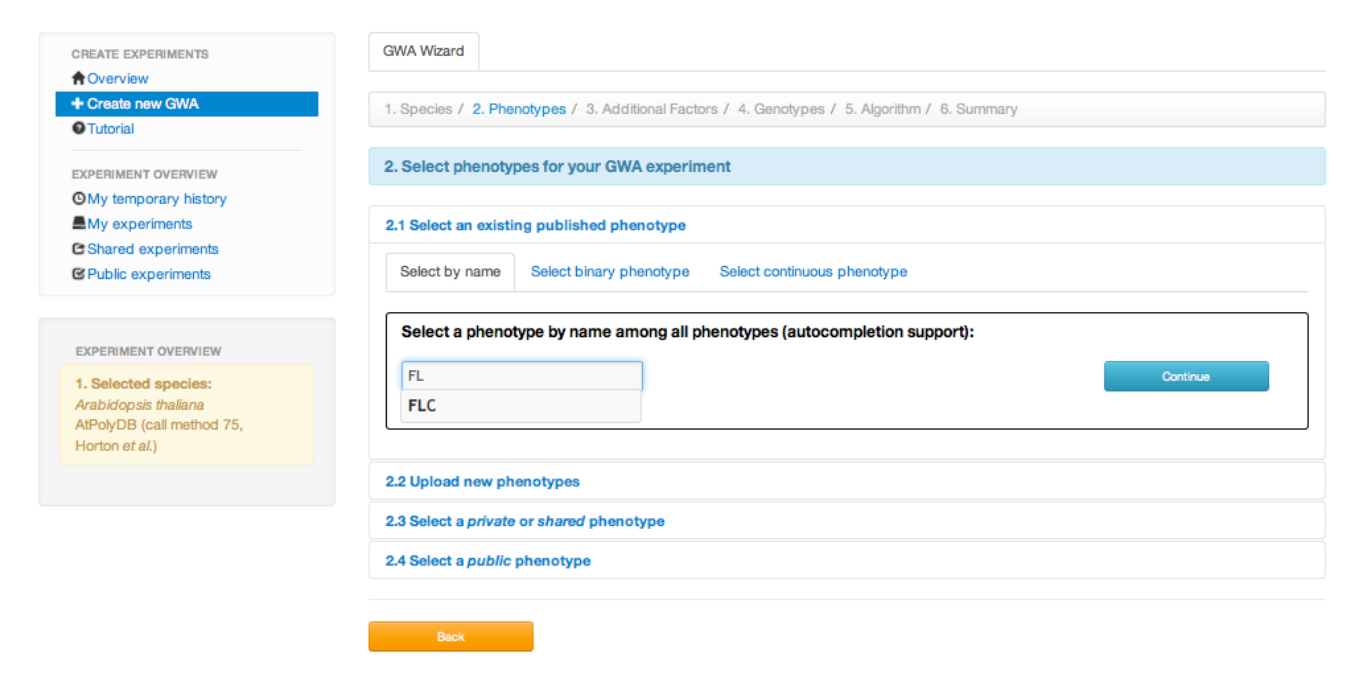} 
		\end{center}
	
	\item In the following step you can add additional factors such as principle components or covariates (e.g. environmental factors, gender specific characteristics). In this tutorial we do not add any additional factors. Click \textbf{Continue} in the tab \textbf{3.1 No additional factors}.
		\begin{center}
		  \includegraphics[]{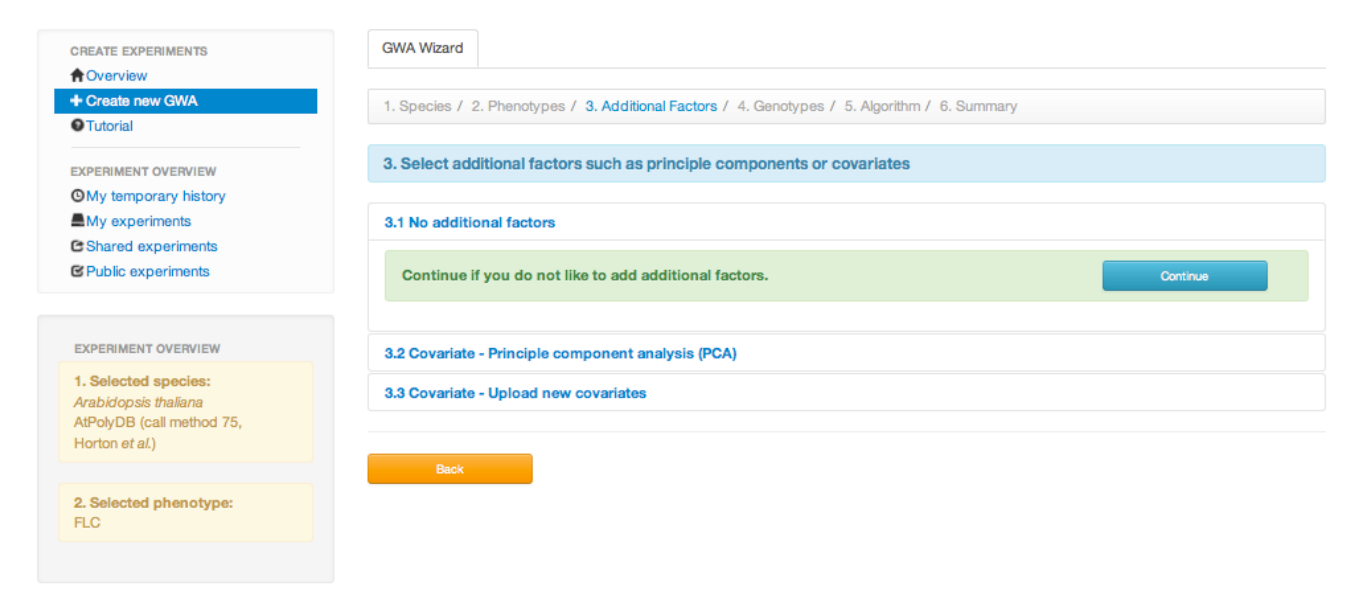} 
		\end{center}
		
	\item Now we have to select genotypic data. You have to choose if you like to use all provided SNPs, specific chromosomes or a region of SNPs. For this tutorial we select chromosome 1 and 5 in the tab \textbf{4.2 Select one or several chromosomes for \textit{Arabidopsis thaliana}} by checking the boxes. Click \textbf{Select chromosomes}.
		\begin{center}
		  \includegraphics[]{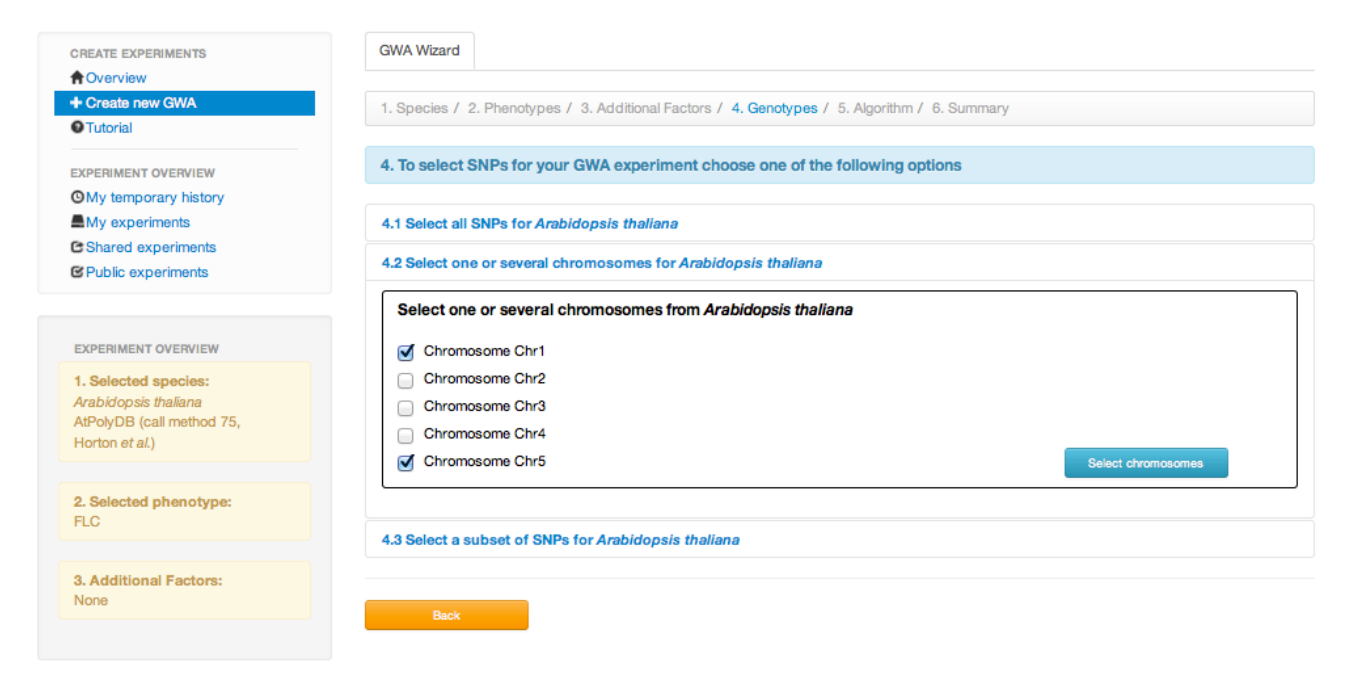} 
		\end{center}
		
	\item To perform a GWAS we have to select a method we intend to use. In the algorithms view one has options to also apply different transformations or filter to the data. The selection of methods and transformations is dependent on the chosen data. Our web application is analyzing the data on the fly and is enabling only those options that are applicable for your data. Here we keep the default settings using a \textbf{Linear Regression} without any transformations. Click \textbf{Continue}.
		\begin{center}
		  \includegraphics[]{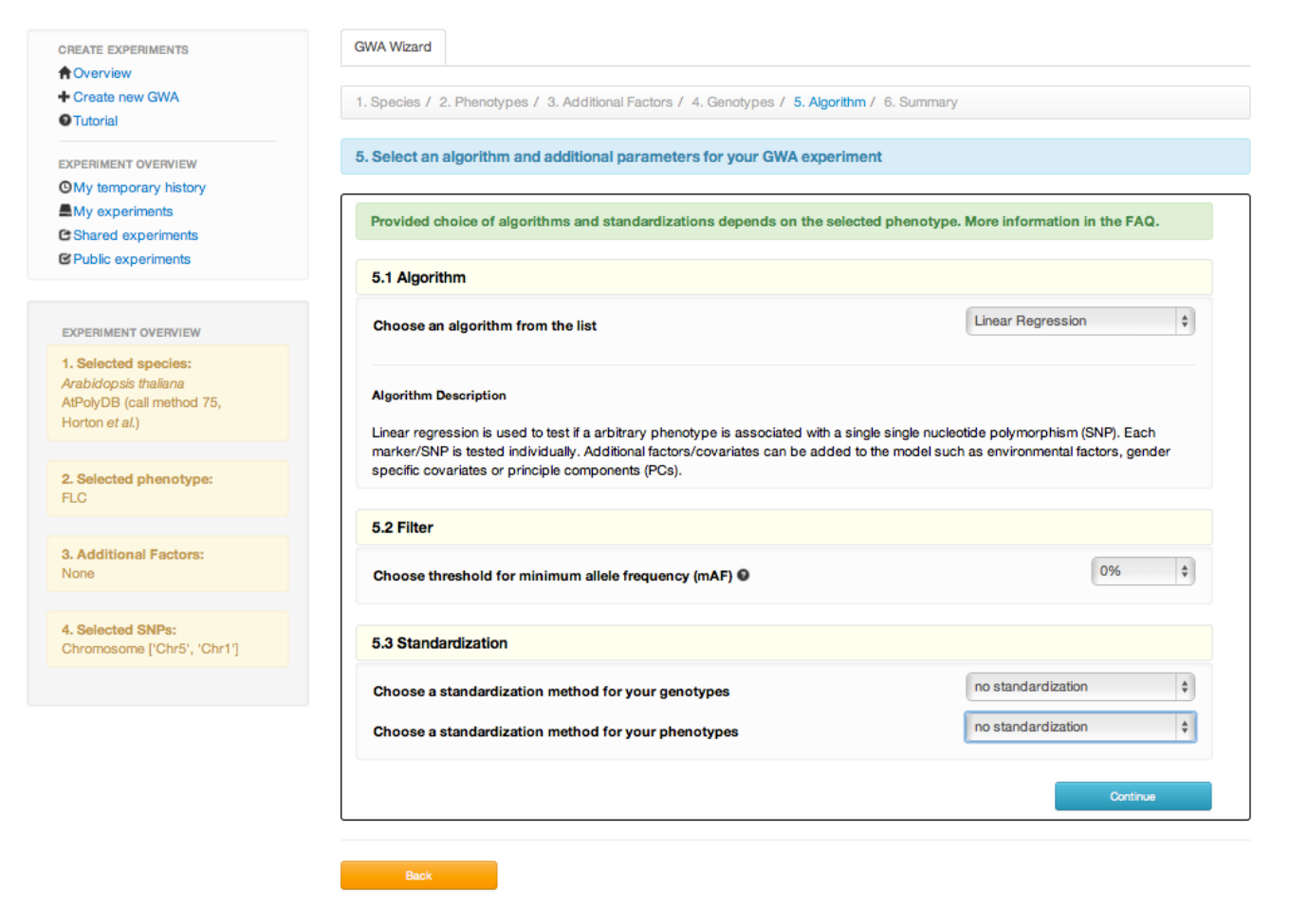} 
		\end{center}
		
	\item In the last step you can check all your inputs again and do necessary adjustments. If everything is correct, you can submit your experiment to the computation servers. For this purpose, simply click \textbf{Submit Experiment}.
	
	\item Finally your experiment is submitted. All computations are running in the background. The current view gets refreshed every 3 seconds. In the meanwhile you could submit new experiments or browse the data. Nevertheless, this experiment is finished in around 60 seconds and you will get automatically redirect to the result view.
	\begin{center}
		  \includegraphics[]{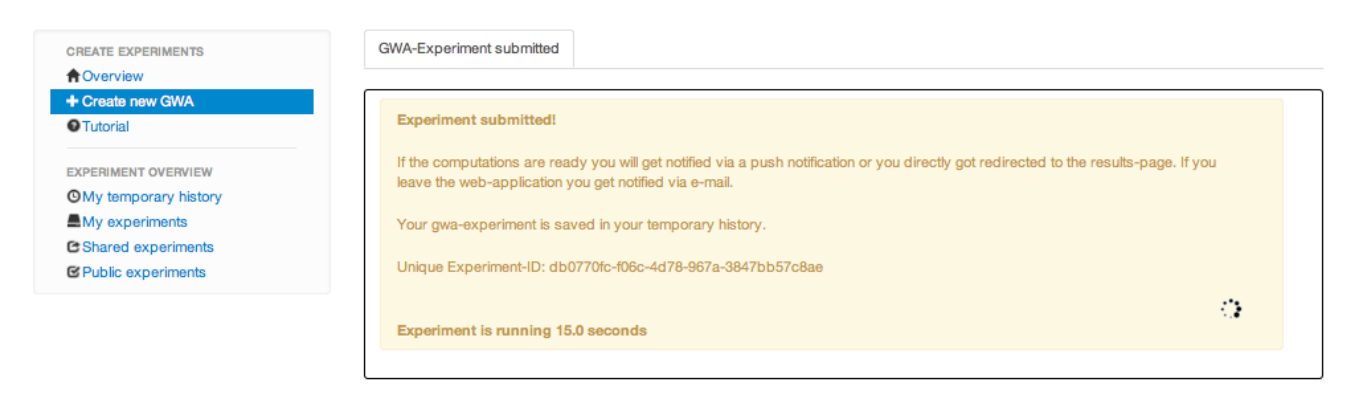} 
		\end{center}
\end{enumerate}

\subsection{How to upload an own phenotype and perform a GWAS on it?}
Here we show how to easily upload new phenotypic data and how to perform a GWAS with it.
\begin{enumerate}
	\item Navigate to the \textit{ \textit{easyGWAS} } wizard\\
		\textbf{Menu: GWA-Experiments $\rightarrow$ Create new GWA}
	
	\item First, select a species and a dataset. Here we choose the species \textit{Arabidopsis thaliana} and the dataset \textit{AtPolyDB (call method 75, Horton et al.)}\cite{horton_genome-wide_2012} and click \textbf{Continue}.
		\begin{center}
		  \includegraphics[]{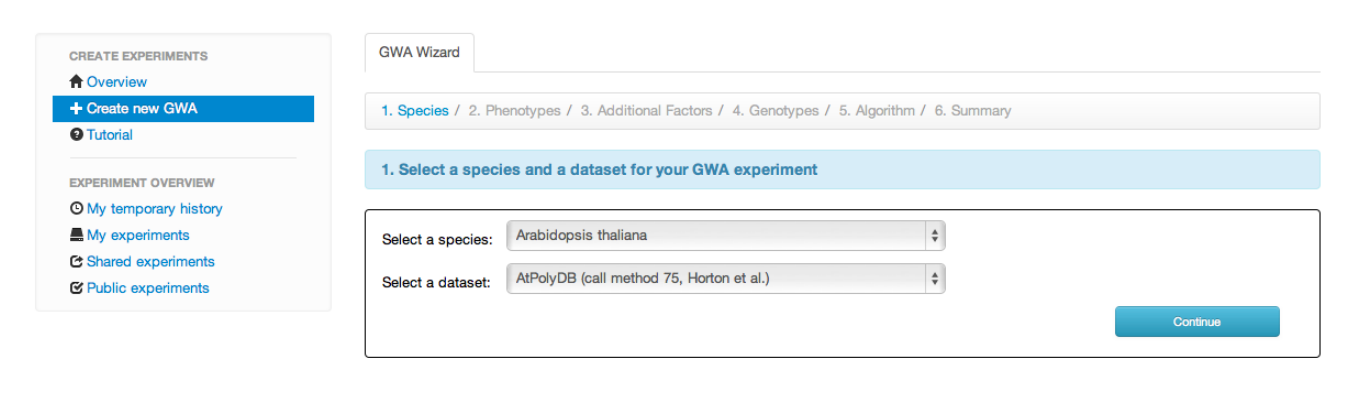} 
		\end{center}	
	
	\item Now navigate to \textbf{2.2 Upload new phenotypic data} and download the linked \textbf{demo} file. Click on \textbf{Choose File} and upload the demo file. Click on \textbf{Continue}.
		\begin{center}
		  \includegraphics[]{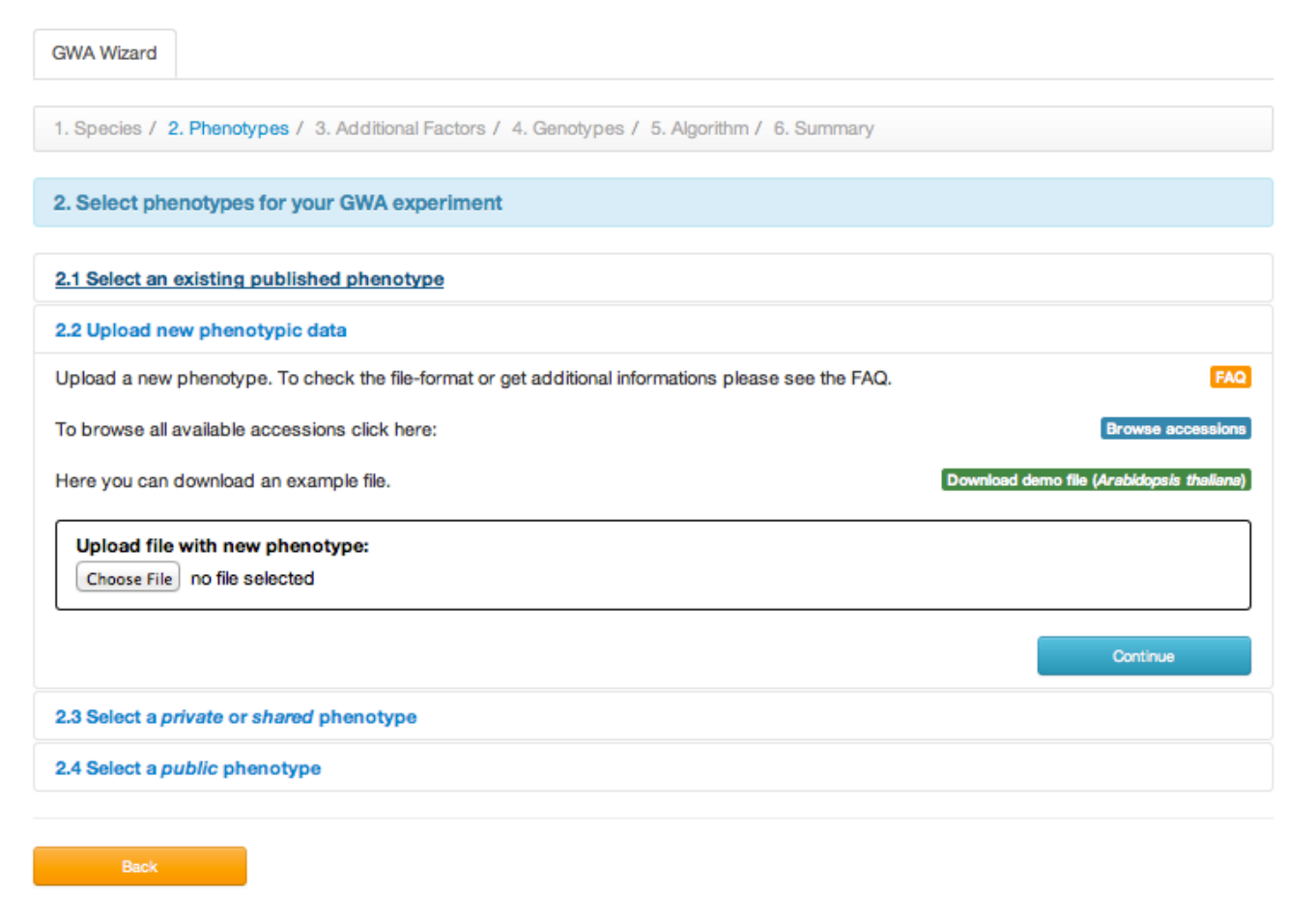} 
		\end{center}	
	
	\item Finally proceed like in Tutorial 5.1. You may try different algorithms and transformations.
		
\end{enumerate}

\subsection{How to store, share or publish your performed GWAS?}

To keep interesting findings and experiments users can save their results permanently using the saving functionality in \textbf{My temporary history}.
Here you can rename your experiment and save it in your experiment history \textbf{My experiments}.
\begin{center}
	\includegraphics[]{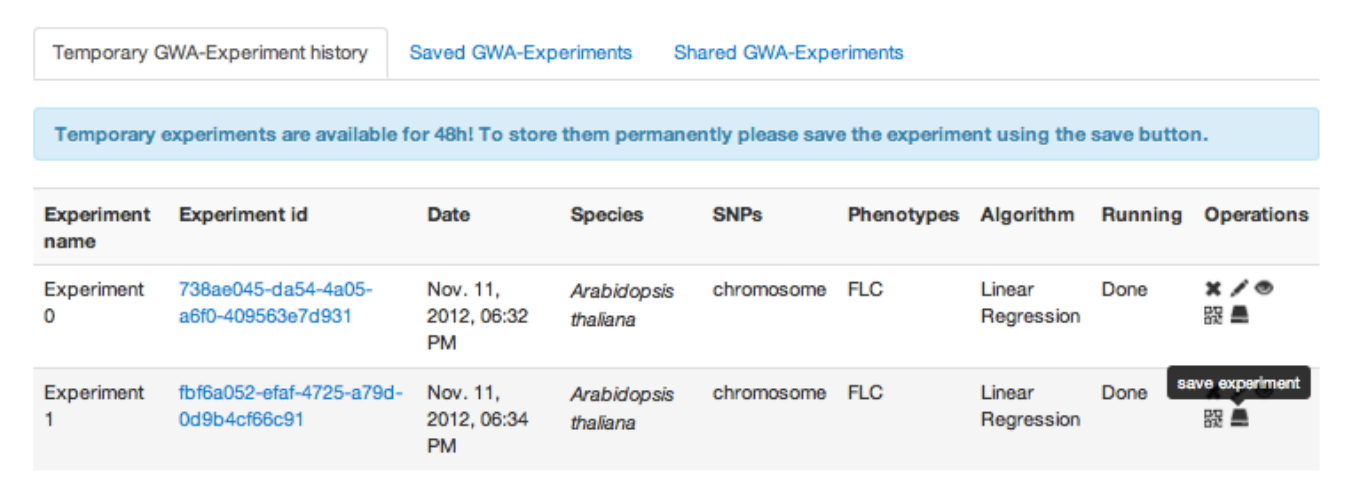} 
\end{center}
To simplify data exchange between colleagues and the scientific community one has the possibility to share and publish saved experiments. For this purpose, open the category \textbf{My experiments} and click on either \textbf{Share Experiment} or \textbf{Publish experiment}. Note that if you like to publish an experiment which was performed on private phenotypes and/or covariates, one has to publish all dependent private data. Please provide meaningful and useful names. 
\begin{center}
	\includegraphics[]{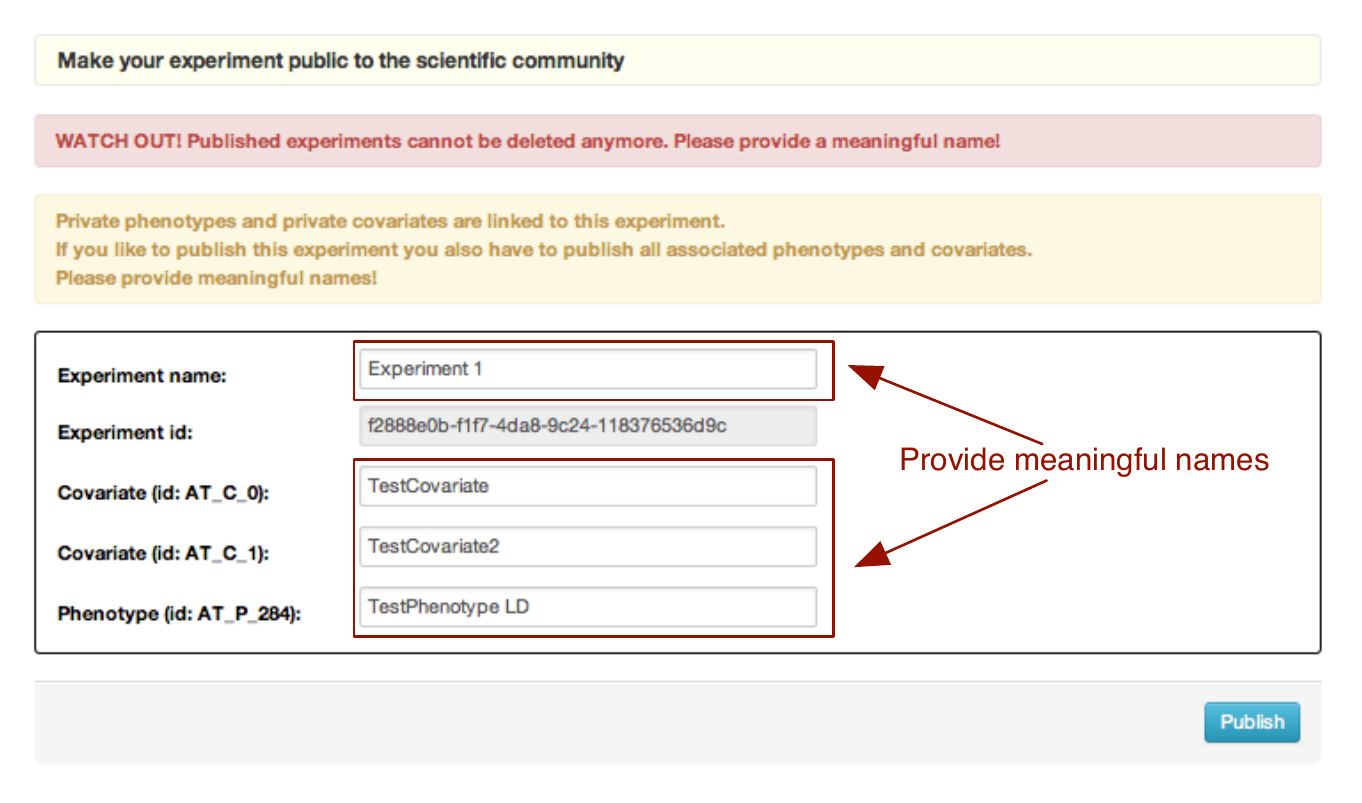} 
\end{center}

\subsection{How to download summary statistics of your results?}
For each experiment summary statistics can be downloaded using the download assistant at the GWAS result page. Click on \textbf{Download Summary Statistics} and choose one of your preferred formats. Right now there are choices for comma-separated files (CSV) and hierarchical data format 5 (HDF5\footref{hdf5}). The summary statistic files contain p-values for each loci and chromosome. The HDF5 file has additional information on which samples were used.
\begin{center}
	\includegraphics[]{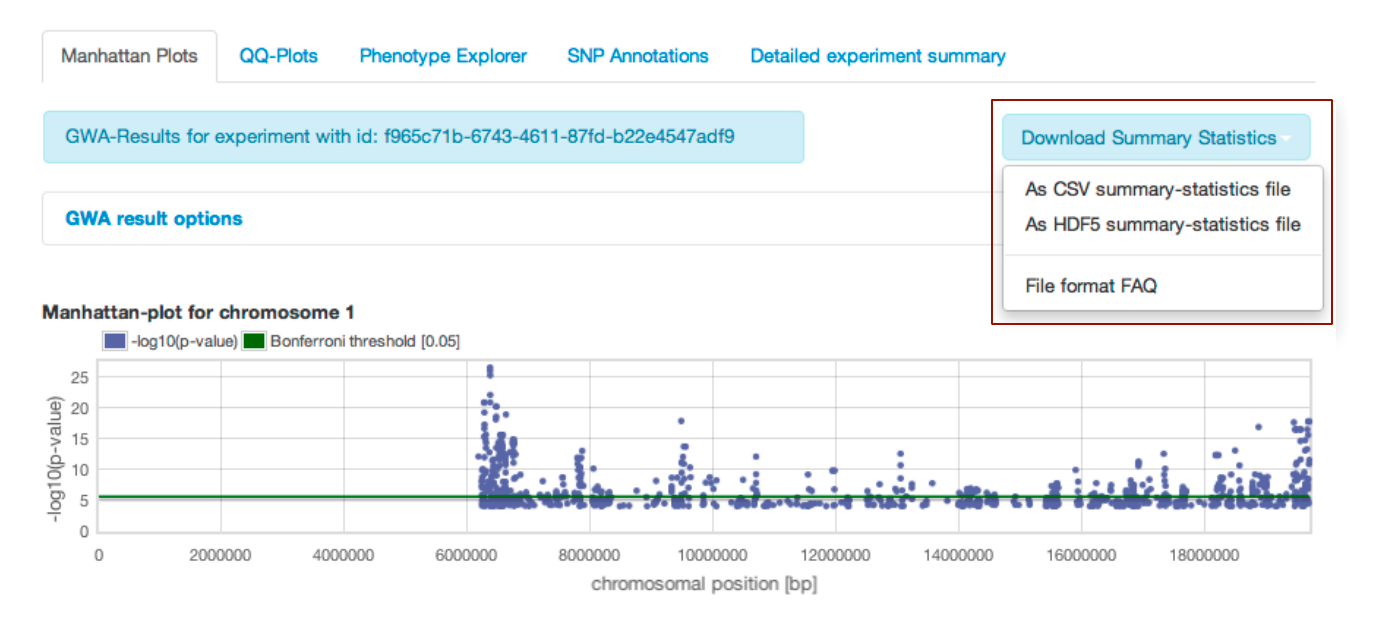} 
\end{center}

\subsection{How to download publicly available data?}
To download whole genotype and phenotype data you can use the \textbf{Download Center} in the main menu. Here you can download all available data for each species and dataset in various formats, supporting PLINK\cite{purcell_plink:_2007}, CSV and HDF5  format.

\bibliographystyle{nature}
\bibliography{references}

\end{document}